# Routing in Wireless Mesh Networks: Two Soft Computing Based Approaches

Sharad Sharma[1,] Shakti Kumar[2] and Brahmjit Singh[3]

[1,3]Deptt. of Electronics & Communication Engineering, National Institute of Technology, Kurukshetra, India
[2]Computational Intelligence (CI) Lab, IST Klawad, Yamunanagar, India

[1]`sharadpr123@rediffmail.com` , [2]`shaktik@gmail.com` , [3]`brahmjit.s@gmail.com`

## ABSTRACT

*Due to dynamic network conditions, routing is the most critical part in WMNs and needs to be optimised. The routing strategies developed for WMNs must be efficient to make it an operationally self configurable network. Thus we need to resort to near shortest path evaluation. This lays down the requirement of some soft computing approaches such that a near shortest path is available in an affordable computing time. This paper proposes a Fuzzy Logic based integrated cost measure in terms of delay, throughput and jitter. Based upon this distance (cost) between two adjacent nodes we evaluate minimal shortest path that updates routing tables. We apply two recent soft computing approaches namely Big Bang Big Crunch (BB-BC) and Biogeography Based Optimization (BBO) approaches to enumerate shortest or near short paths. BB-BC theory is related with the evolution of the universe whereas BBO is inspired by dynamical equilibrium in the number of species on an island. Both the algorithms have low computational time and high convergence speed. Simulation results show that the proposed routing algorithms find the optimal shortest path taking into account three most important parameters of network dynamics. It has been further observed that for the shortest path problem BB-BC outperforms BBO in terms of speed and percent error between the evaluated minimal path and the actual shortest path.*

## KEYWORDS

*Wireless Mesh Network, Cost function, Fuzzy Logic, Big Bang Big Crunch, Biogeography Based Optimization*

## 1. INTRODUCTION

Wireless Mesh Networks are rapidly deployable, dynamically self organizing, self configuring, self healing, self balancing and self aware multi hop networks. In these networks each node (stationary or mobile) has the capability to join and create a network automatically by sensing nodes with a similar capability within its radio range. WMNs can be categorized to three types: (1) Infrastructure Mesh; (2) Client Mesh and (3) Hybrid Mesh [1].

The routing algorithms of a WMN must work in a decentralized, self-organizing and self configuring manner. The routing protocols developed for MANETs can usually be applied to WMNs as well e.g. Ad hoc On-demand Distance Vector (AODV) [2], Topology Broadcast based Reverse Path Forwarding (TBRPF) [3], Dynamic Source Routing (DSR) [4] etc. The existing MAC and routing protocols applied to WMNs do not provide enough scalability [5]. The factors like resource allocation, interference avoidance and rate adaptation across multiple hops critically affects the routing path selection [6]. In a WMN the performance parameters can be categorized

International Journal of Mobile Network Communications & Telematics ( IJMNCT) Vol. 3, No.3, June 2013as per flow; per node; per link; inter flow and network wide parameters. These routing metrics are Airtime Cost Routing Metric [6], Hop Count, Per-Hop Round Trip Time (RTT) [7], Metric of Interference and Channel-Switching (MIC) [8], Expected Transmission Count (ETX) [9], Expected Transmission on a Path (ETOP) [10], Expected Transmission Time (ETT) and Weighted Cumulative ETT (WCETT) [11], Low Overhead Routing Metric [12, 13], Effective Number of Transmissions (ENT) [14], Bottleneck Link Capacity (BLC) [15], Per-Hop Packet Pair Delay and Expected Data Rate (EDR) [16], etc. The comparison of performance metrics on a routing algorithm is discussed in [16]. Soft computing based techniques provide the optimal solution and quickly adapts to dynamic environmental changes [17]. In this paper, we propose two soft computing approaches for addressing the routing problem in WMNs, namely BB-BC and BBO.

This paper is organized into five sections. Section I presents the motivation for the present work. Section II introduces the Big Bang Big Crunch (BB-BC) and Biogeography Based Optimization (BBO) algorithms. In section III we present a fuzzy logic based approach to evaluate integrated link cost (ILC) for each link consisting of throughput, end-to-end delay and average jitter of a given path. In section IV, BB-BC and BBO are employed to evaluate the shortest path in the configured network. The section V compares the performance of two algorithms. Conclusions are drawn in Section VI.

## 2. BIG BANG BIG CRUNCH (BB-BC) AND BIOGEOGRAPHY BASED OPTIMIZATION (BBO)

### 2.1 BB-BC Optimization Algorithm

The Big Bang theory is one of the most widely accepted theories of the evolution of this universe [18]. The BB-BC theory believes that energy discharged by the initial explosion i.e., kinetic energy, is counterbalanced by the energy of bodies attraction known as gravitational pull. If there is enough mass so that the later is bigger than the first when a critic density is reached, the expansion will stop and the universe will start to contract, leading to an end very similar to its beginning, named by the scientists as the Big Crunch (Great Implosion). In the Big Bang phase, energy dissipation produces disorder and randomness as the main feature of this phase. In the Big Crunch phase, randomly distributed particles are drawn into an order. This theory of repeated big bang followed by big crunch phases forms the basis of an optimization algorithm called the Big Bang-Big Crunch optimization algorithm [19, 20].

Primarily a set of candidate solutions (population) is generated randomly in the search space. The fitness as defined by the objective function, of each solution is enumerated and ranked accordingly. After the random Big Bang phase contraction is applied in Big Crunch phase to compute the centre of mass as:

$$x_c = \frac{\sum_{i=1}^{N} \frac{1}{f^i} x_i}{\sum_{i=1}^{N} \frac{1}{f^i}} \qquad (1)$$

where $x_c$ = position of the centre of mass; $x_i$ = position of ith candidate; $f^i$ = fitness function value of candidate $i$; N = population size.

Best fit individual can also be considered as the centre of mass instead of using Equation (1) alternatively. Generate new population around the centre of mass by adding or subtracting a normal random number whose value decreases as the iterations elapse. This can be formalized as

$$x^{new} = x_c + lr/k \qquad (2)$$

where $x_c$ stands for center of mass, $l$ is the upper limit of the parameter, $r$ is a normal random number and $k$ is the iteration step. Then new point $x^{new}$ is upper and lower bounded.

Pseudo Code for optimal path evaluation in WMNs using BB-BC based Algorithm is shown in Figure 1.

```
begin
    / BB-BC Parameter Initialization for WMN/
    Define Number and location of the nodes Source Node, Terminal Node, Number of
    Paths, Number of Iterations,
    / End of BB-BC Parameter Initialization/
    /Building of paths and Connectivity Matrix/
    for i = 1 : n                                          / all n Nodes /
        for j = 1 : n                                      / all n Nodes /
            if distance (i, j) <= R (radio range of a node)
                connectivity_matrix(i, j) = 1              /routing table maintenance/
                Integrated_Link_Cost (i, j )= f (Throughput, Delay, Jitter)
                /Integrated Link Cost Evaluation using Fuzzy System/
            end if
        end for j
    end for i
    / Build paths between source and terminal node /
    while (t < MaxGeneration or Termination Criteria not met)
    Randomly generate initial population of k paths         /Big Bang Phase/
    Compute the ILC of all the candidate solutions
    Sort the population from best to worst based on ILC   /No.1 path is the Optimal path/
    Compute the center of mass x_c                          /Big Crunch Phase/
    Generate new candidate solutions around x_c by adding or subtracting a normal-
        -random Number
    end while
    wait for stipulated time/ wait for an event

Postprocess results and visualization;
end
```

Figure 1: Pseudo Code of BB-BC based Algorithm for optimal path evaluation in WMNs

## 2.2 BBO ALGORITHM

BBO is the study of how species are articulated on the landscape in space and time. MacArthur and Wilson (1963) [21] first suggested that the number of species of a given taxon that become established on an island represents a dynamic equilibrium controlled by the rate of immigration of new species and the rate of extinction (emigration) of previously established species. Based upon the dynamical equilibrium theory Dan Simons [22] proposed BBO algorithm. Since its first application, this meta-heuristic approach has been applied successfully to some engineering applications. We apply this algorithm to evaluate minimal cost path.

*BBO Algorithm* [23, 24]:
- Initialize the parameters of BBO:
  - maximum species count $S_{max}$ and species count probability $P_S$ of each habitat, maximum migration rates $E$ and $I$, immigration rate $\lambda$, emigration rate $\mu$, the maximum mutation rate $m_{max}$, elitism parameter and habitat modification probability
- Initialize a random set of habitats, each habitat corresponding to a potential solution to the given problem.
- Do while not end of termination criteria (number of iterations)
- Compute "fitness" (HSI-habitat suitability index) for each habitat (solution). The variables, characterizes habitability are Suitability Index Variables (SIVs).
- For each habitat, map the HSI to the number of species S. (Species Count is inversely proportional to Cost)
- Compute $\lambda$ and $\mu$ for each solution.

$$\lambda_k = I \left( 1 - \frac{k}{n} \right) \quad (3)$$

$$\mu_k = E \left( \frac{k}{n} \right) \quad (4)$$

here the parameters are given as-
I- maximum possible immigration rate, E- maximum possible emigration rate, k- Number of species of $k^{th}$ individual and n- maximum number of species
- Modify each no elite habitat probabilistically using immigration and emigration rates & recompute each HSI
- For each habitat, update the probability of its species count.

$$\dot{P} = \begin{cases} \lambda_0 P_0 + \mu_1 P_1, & k = 0 \\ -(\lambda_k + \mu_k)P_k + \lambda_{k-1}P_{k-1} + \mu_{k+1}P_{k+1}, & 1 \ll k \ll n-1 \\ -\mu_n P_n + \lambda_{n-1}P_{n-1}, & k = n \end{cases} \quad (5)$$

- Mutate each non-elite habitat based on its probability.
- If acceptable solution (optimal path) has been found then stop
- End

Pseudo code for optimum path selection in WMNs using BBO is presented in figure 2.

**begin**
    **/ BBO Parameter Initialization for WMN/**
    *Define Number and location of the nodes, Source Node, Terminal Node, Number of Paths, Number of Iterations, $S_{max}$, $m_{max}$, E, I, $\lambda$ and $\mu$*
    **/Building of paths and Connectivity Matrix/**
    **for** *i = 1 : n*                                             / all n Nodes /
        **for** *j = 1 : n*                                        / all n Nodes /
            **if** *distance (i, j) <= R (radio range of a node)*
    *connectivity_matrix(i, j) = 1*                        /routing table maintenance/
    *Integrated_Link_Cost (i, j )= f (Throughput, Delay, Jitter)*
    */ILC Evaluation using Fuzzy System/*
        **end** *if*
        **end for** *j*
    **end for** *i*
    **/** *Build paths between source and terminal node* **/**
    **while** *(t < MaxGeneration or Termination Criteria not met)*
    *Randomly generate initial population of k paths (Habitats)*     */Initial Population/*
    *Compute the fitness (HSI) of all the candidate solutions (ILC)*

```
        Sort the population from best to worst based on ILC   /No.1 path is the Optimal path/
        Map the HSI to the no. of species of each individual
        Compute the immigration rate λi and the emigration rate μi for each individual
        Generate new candidate solutions and Update the probability
       end while
    wait for stipulated time/ wait for an event
Postprocess results and visualization;
end
```

Figure 2: Pseudo Code of BBO based Algorithm for optimal path evaluation in WMNs

## 3. System Model

To analyse and optimize the performance of routing algorithm of WMN environment simulations were performed for a pre defined scenario in QualNet Simulator [25]. In this simulation, a network of 25 nodes is considered that are placed within a 1500m X 1500m area and operating over 300 seconds. A two ray ground propagation model is used with log normal shadow fading. The transmission power of the nodes is set to 15dBm and the transmission range of the nodes is 250 meters. The data transmission rate is 2Mbits/sec. At the physical layer 802.11b and at MAC layer MAC 802.11 is used. The traffic source is implemented using Constant Bit Rate (CBR), sending at the rate of 1 packet/sec.

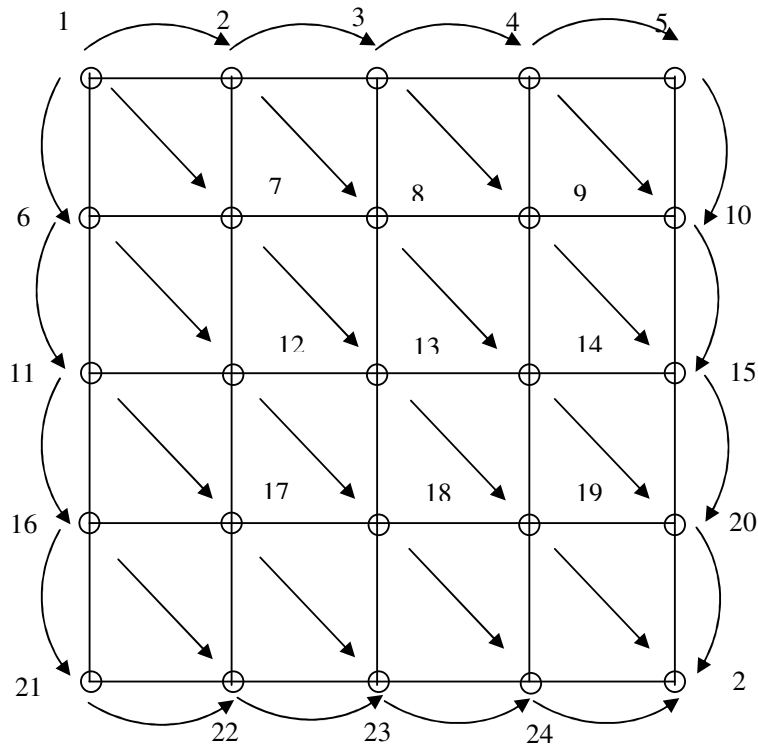

Figure: 3 Scenario of a 25 node static Wireless Mesh Network

## 4. Path Cost Evaluation

For our model we considered 25, 64, 100 and 2500 nodes in this network. First node is assumed to transmit data packets to the last node. To decide which path or route is to be used for any type of traffic or condition depends on the current values of parameters at the nodes or the links.

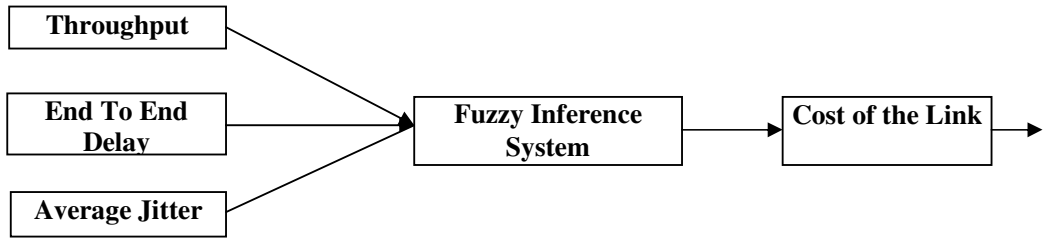

Figure 4:   Fuzzy System Model for Integrated Link Cost Evaluation

In the proposed work a routing metric of integrated throughput, end to end delay and jitter has been considered at each node for all of its neighbouring nodes. Considering only forward links a cost function has been proposed using fuzzy logic [26]. For our model to work we assume that each node contains a software based fuzzy logic cost evaluation system as given in figure 4.

## 5. Results and Discussion

Two algorithms BB-BC and BBO were implemented in MATLAB v 7.6.0 (R2008a). Numerical results were computed for both algorithms and are given in Table 1. It is observed from figure 5 and Table 1 that for a 25 node network, the minimum path cost is evaluated with zero error. Further only 30 iterations suffice to make a quick decision about the shortest/near shortest available path to ensure the optimal performance of the model. For 64 node WMN, initially the BB-BC performs better (table 1). However as the number of iterations is increased from 30 to 100, the error of the two reduces to zero. Thus both again produced shortest path for the given scenario. But time taken for finding the shortest path for BBO was 4.96 seconds as against 0.29 seconds for BB-BC. For a 100-node network, we found that minimum error achievable by BBO after 100 iterations was 2.38% as compared to 1.36% of BB-BC. The time taken by BBO for 100 iterations was 9.55 seconds against 0.58 seconds in case of BB-BC. As the number of nodes is increased to 2500, BB-BC took 21.26 seconds for 100 iterations to enumerate a shortest path with 1.90% error. BBO in contrast took 1 hour 12 minutes 29 seconds for the same number of iterations and produced 2.982% error.

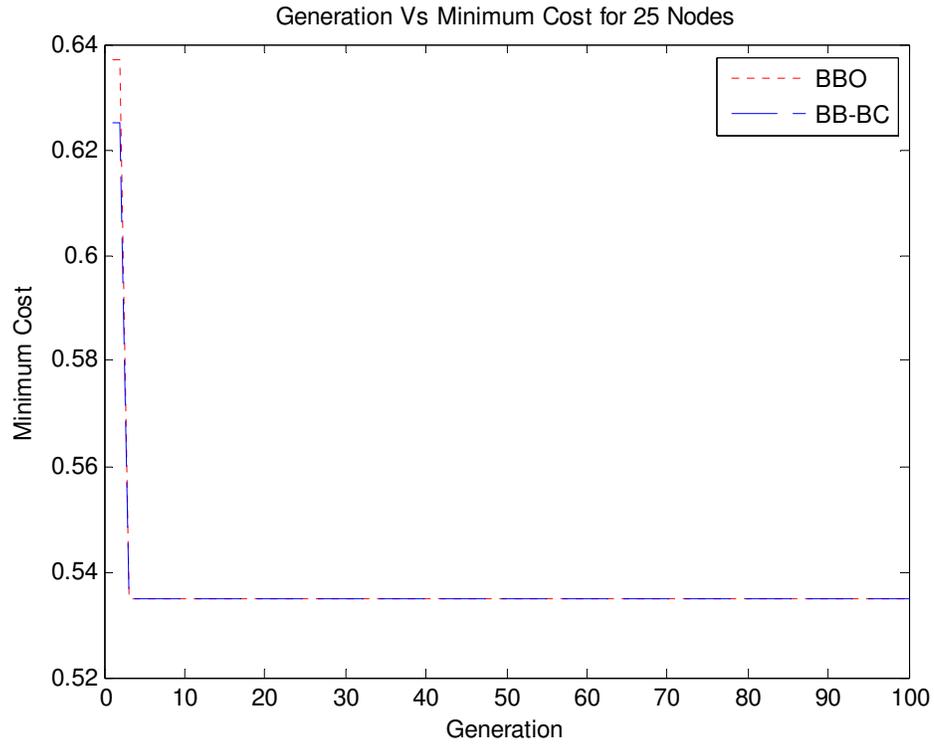

Figure 5: Generation vs Minimum Cost for 25 Node network for 100 generations

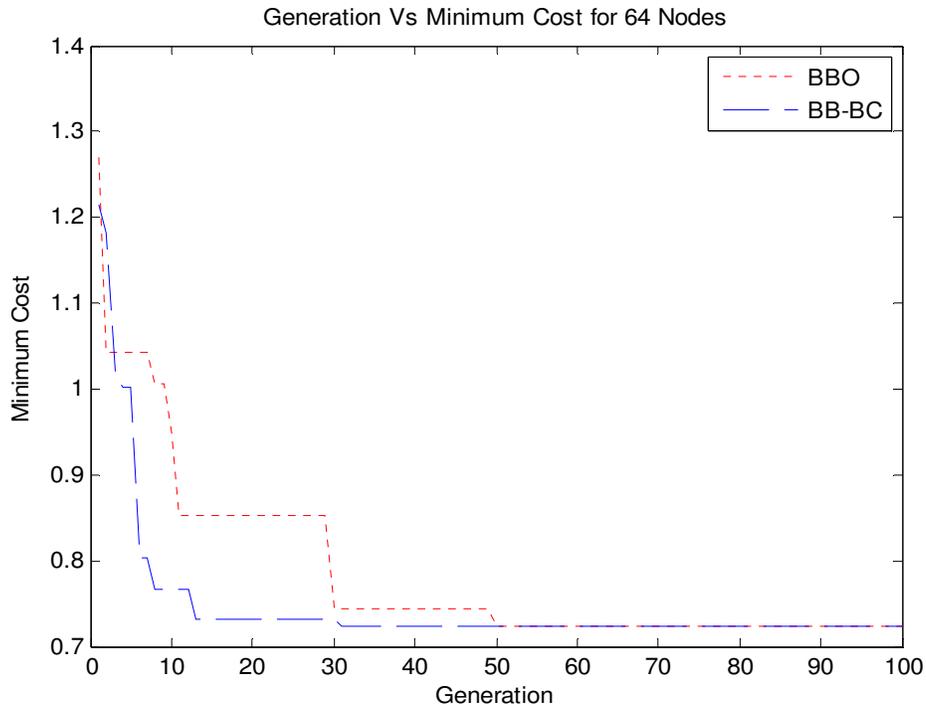

Figure 6: Generation vs Minimum Cost of 64 Node network for 100 generations

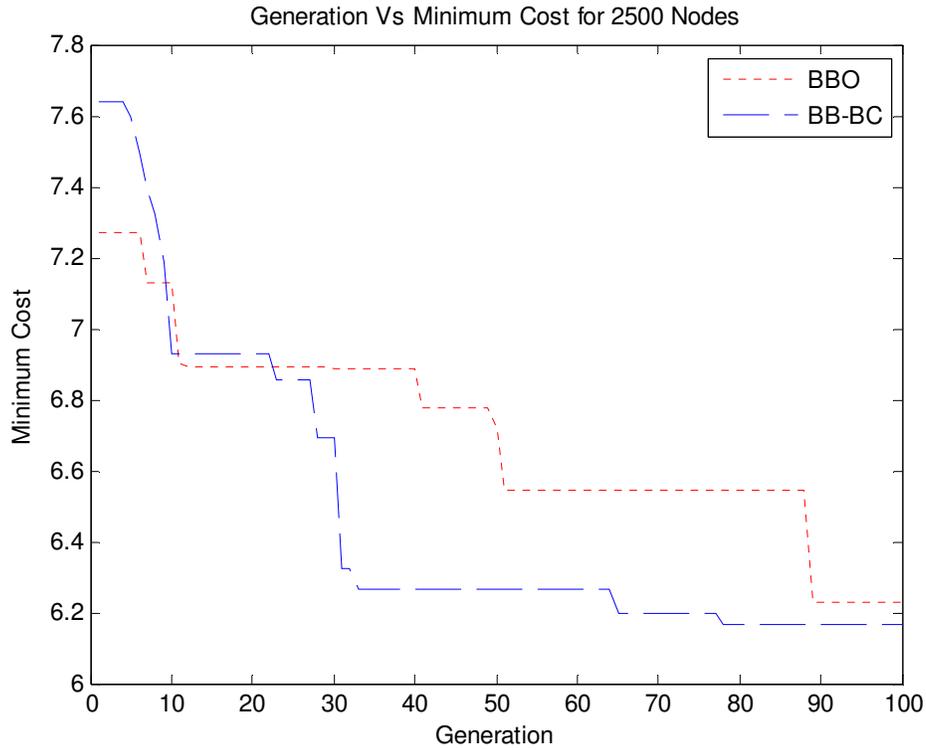

Figure 7: Generation vs Minimum Cost of 2500 Node network for 100 generations

Table 1: Results of BB-BC and BBO algorithms

| No. of Nodes | No. Of Generations | BB-BC | | | BBO | | |
|---|---|---|---|---|---|---|---|
| | | Path Cost | % Error | Time (sec) | Path Cost | % Error | Time (sec) |
| 25 | 30 | 0.5350 | 0.00 | 0.104528 | 0.5350 | 0.00 | 1.813251 |
| | 50 | 0.5350 | 0.00 | 0.139245 | 0.5350 | 0.00 | 2.194382 |
| | 100 | 0.5350 | 0.00 | 0.214183 | 0.5350 | 0.00 | 2.584156 |
| 64 | 30 | 0.7315 | 1.147 | 0.224916 | 0.7442 | 2.903 | 2.761474 |
| | 50 | 0.7232 | 0.00 | 0.271387 | 0.7242 | 0.138 | 3.420020 |
| | 100 | 0.7232 | 0.00 | 0.289708 | 0.7232 | 0.00 | 4.960113 |
| 100 | 30 | 0.9197 | 6.05 | 0.325719 | 0.9274 | 6.94 | 3.977115 |
| | 50 | 0.8990 | 3.66 | 0.407832 | 0.9135 | 5.33 | 6.76877 |
| | 100 | 0.8790 | 1.36 | 0.581593 | 0.8879 | 2.38 | 9.557044 |
| 2500 | 30 | 6.3266 | 4.587 | 7.450026 | 6.7582 | 11.722 | 1278.180244 |
| | 50 | 6.2640 | 3.55 | 11.174060 | 6.6384 | 9.741 | 1990.837452 |
| | 100 | 6.1642 | 1.90 | 21.268608 | 6.2295 | 2.982 | 4249.000040 |

International Journal of Mobile Network Communications & Telematics ( IJMNCT) Vol. 3, No.3, June 2013

## 6. Conclusion and Future Scope of the Work

As the network complexity grows, time required to evaluate the exact shortest path increases. This paper presented two recent soft computing based approaches to evaluate near shortest path in a WMN. WMNs being highly dynamic need shortest path periodically to modify their routing strategies. Further the shortest path evaluation technique must be able to compute a near shortest path as quickly as possible so as to keep the routing performance optimal. Keeping the routing policies in mind we proposed an integrated path length measure that takes into account throughput, end-to-end delay and jitter of the link. A fuzzy based approach was used for computing the distance between two adjacent nodes in terms of this integrated metric. Using this integrated cost function as the path length metric we evaluated shortest path between source and destination node using BB-BC and BBO approaches.

A large number of experiments were conducted to find the shortest path in a 25, 64, 100 and 2500 node models. It was observed that BB-BC approach clearly outperformed BBO approach in terms of evaluation time and error between the actual shortest path and the evaluated near shortest path, evaluated by these two approaches. For the small WMNs of 25, 64 and 100 nodes though the performance of the two was comparable with BB-BC always producing better results. For a 2500 node model BB-BC produced its shortest path with 1.90% error in 21.26 seconds after 100 iterations whereas BBO generated its best path with 2.982% error in 1 hr 12 minute and 29 seconds. Thus, clearly establishing the superiority of BB-BC over BBO for finding the optimal path in a WMN.

**Authors**

**Sharad Sharma,** received his B.Tech in Electronics Engineering from Nagpur University, Nagpur, India in 1998 and M.Tech in Electronics and Communication Engineering from Thapar Institute of Engineering and Technology, Patiala, India in 2004. He has a teaching experience of 14 years. He has conducted many workshops on Soft Computing and its applications in engineering, Wireless Networks, Simulators etc. He has a keen interest in teaching and implementing the latest techniques related to wireless and mobile 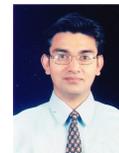 communications. He opened up a student chapter of IEEE as Branch Counselor. Presently, he is working towards the Ph.D. degree in Electronics and Communication Engineering from National Institute of Technology, Kurukshetra, India. His research interests are routing protocol design, performance evaluation and optimization for wireless mesh networks using nature inspired computing.

**Prof.(Dr.)Shakti Kumar** received his M.S.(Electronics and Control) from Birla Institute of Technology & Science (BITS), Pilani and Ph.D.(Electronics and Computer Engg.) from National Institute of Technology (Formerly REC), Kurukshetra. He has served as a faculty member in the department of Electronics & Computer Engineering, NIT Kurukshetra, Department of Computer Engineering, Atlim University, Ankara, Turkey and Alghurair University, Dubai, UAE. He was Co-ordinator, Centre for Excellence in Reliability 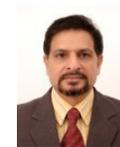 Engineering at REC (Now National Institute of Technology), Kurukshetra. He has published more than 165 research articles in various International/National journals and conferences, 7 book chapters and has published one patent. He has chaired 13 International/National Conferences. He is member of various professional bodies such as IEEE and ISTE. Dr. Kumar was jointly awarded **"Sir Thomas Ward Memorial"** Subject Award by Institution of Engineers (India) in December 2006.

**Dr. Brahmjit Singh,** received his B.E. Degree in Electronics Engineering from Malaviya National Institute of Technology (then MREC), Jaipur in 1988, M.E. with specialization in Microwave and Radar from Indian Institute of Technology, Roorkee (then UoR) in 1995 and Ph.D. from Guru Govind Singh Indraprastha University, Delhi in 2005. He is with Electronics and Communication Engineering Deptt., National Institute of Technology working as Professor since 2006. He has published 75 research papers in International/National journals and conferences. His current research interests include mobility management in mobile cellular networks, security in wireless networks, wireless sensor networks and cognitive radio. He has been awarded **"Sir Thomas Ward Memorial"** Subject Award for the paper titled, "*Optimizing Handover Performance in Microcellular Systems*" on behalf of Institution of Engineers (India) in December, 2006. He is the member of IEEE, IETE and ISTE. 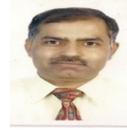